\documentclass[twocolumn, 10pt,english,aps,prb,superscriptaddress,bibnotes,amsmath,amssymb,floatfix]{revtex4-1}
\usepackage[colorlinks=true,citecolor=blue,linkcolor=magenta]{hyperref}
\usepackage[utf8]{inputenc}
\usepackage[english]{babel}
\usepackage[T1]{fontenc}
\usepackage{textcomp}
\usepackage{soul}
\usepackage{url}
\usepackage{makecell}
\usepackage{graphicx}
\usepackage{epstopdf}

\usepackage[]{changes}
\setaddedmarkup{\textcolor{red}{#1}}
\setdeletedmarkup{\textcolor{blue}{\sout{#1}}}

\graphicspath{{./figures/}}
\makeatletter
 
\newcommand{\Rmnum}[1]{\expandafter\@slowromancap\romannumeral #1@} 
\makeatother
\usepackage{multirow}

\begin{document}
\title{Filter-Free Indistinguishable Photon Generation from\\Continuous-Wave-Driven Integrated Microresonators}

\author{Ruiyang Chen}
\affiliation{School of Physical Sciences and Hefei National Laboratory, University of Science and Technology of China, Hefei 230026, China}
\affiliation{International Quantum Academy and Shenzhen Futian SUSTech Institute for Quantum Technology and Engineering, Shenzhen 518048, China}

\author{Sicheng Zeng}
\affiliation{International Quantum Academy and Shenzhen Futian SUSTech Institute for Quantum Technology and Engineering, Shenzhen 518048, China}
\affiliation{Shenzhen Institute for Quantum Science and Engineering, Southern University of Science and Technology, Shenzhen 518055, China}

\author{Yuan Chen}
\affiliation{International Quantum Academy and Shenzhen Futian SUSTech Institute for Quantum Technology and Engineering, Shenzhen 518048, China}

\author{Sanli Huang}
\affiliation{School of Physical Sciences and Hefei National Laboratory, University of Science and Technology of China, Hefei 230026, China}
\affiliation{International Quantum Academy and Shenzhen Futian SUSTech Institute for Quantum Technology and Engineering, Shenzhen 518048, China}

\author{Zeying Zhong}
\affiliation{International Quantum Academy and Shenzhen Futian SUSTech Institute for Quantum Technology and Engineering, Shenzhen 518048, China}
\affiliation{Shenzhen Institute for Quantum Science and Engineering, Southern University of Science and Technology, Shenzhen 518055, China}

\author{Zhen Chen}
\affiliation{School of Physical Sciences and Hefei National Laboratory, University of Science and Technology of China, Hefei 230026, China}
\affiliation{International Quantum Academy and Shenzhen Futian SUSTech Institute for Quantum Technology and Engineering, Shenzhen 518048, China}

\author{Xue Bai}
\affiliation{International Quantum Academy and Shenzhen Futian SUSTech Institute for Quantum Technology and Engineering, Shenzhen 518048, China}
\affiliation{Qaleido Photonics, Shenzhen 518048, China}

\author{Yi-Han Luo}
\email{luoyh@iqasz.cn}
\affiliation{International Quantum Academy and Shenzhen Futian SUSTech Institute for Quantum Technology and Engineering, Shenzhen 518048, China}

\author{Junqiu Liu}
\affiliation{School of Physical Sciences and Hefei National Laboratory, University of Science and Technology of China, Hefei 230026, China}
\affiliation{International Quantum Academy and Shenzhen Futian SUSTech Institute for Quantum Technology and Engineering, Shenzhen 518048, China}

\begin{abstract}
Quantum networks require scalable photon sources combining narrow linewidth, high efficiency, and high indistinguishability.
Microresonator photon-pair sources are promising candidates, yet a source-only description based on the joint spectral amplitude (JSA) predicts near-zero Hong--Ou--Mandel (HOM) interference visibility between photons generated by independent, CW-driven microresonators.
In this work, we show that the observed HOM interference is not determined by the JSA alone.
By combining finite-time detection with cavity-enhanced spontaneous four-wave mixing, we characterize a detector-conditioned heralded state governed by the idler-photon detection window. 
We further demonstrate that independently optimizing the idler and signal detection windows allows both heralded-photon indistinguishability and intrinsic heralding efficiency to approach unity, without spectral filtering or complex source engineering.
Utilizing integrated high-$Q$ silicon nitride microresonators, we achieve HOM visibilities of 0.992(8) and 0.942(12), without background subtraction, at fourfold count rates of 4.5(3) and 12.2(6)~Hz, respectively. 
Our work establishes CW-driven high-$Q$ microresonators as a robust and scalable platform for quantum-network primitives.
\end{abstract}

\maketitle


Photons serve as an ideal platform for quantum networks \cite{Kimble:08}, acting as flying qubits that are robust against environmental decoherence \cite{Pan:12}.
Fundamental protocols, including quantum teleportation \cite{Bennett:93, Bouwmeester:97} and entanglement swapping \cite{Pan:98}, require high-visibility Hong--Ou--Mandel (HOM) interference \cite{Hong:87} between photons from independent sources. 
Practical large-scale deployment demands photon sources that simultaneously feature narrow linewidths, high efficiency, near-unity indistinguishability, and a clear path toward scalable integration.

Microresonator photon-pair source emerges as a premier candidate for these requirements, generating bright, narrowband photon pairs in compact footprints across diverse material platforms \cite{Ma:17, Fan:23, Chen:2024, Li:25, Zeng:24, Steiner:21, Ma:20, Pang:25, Luo:26}.
Their narrow linewidths preserve indistinguishability over long-distance fiber transmission and enable efficient light--matter interfaces.
Consequently, they have already been deployed as critical building blocks for quantum teleportation \cite{Llewellyn:20}, entanglement swapping \cite{Samara:21}, and photon fusion \cite{Alexander:25}.

Continuous-wave (CW) pumping \cite{Samara:21} is particularly advantageous for field-deployable systems, as it permits the use of chip-scale integrated lasers \cite{Xiang:21, Sun:25}.
However, from the conventional perspective, the joint spectral amplitude (JSA) dictates a spectral purity that bounds the HOM visibility between independent heralded photons~\cite{Law:00}. 
Specifically, CW pumping imposes stringent energy conservation that results in strong signal--idler frequency correlations, leading to a non-factorable JSA and hence a vanishingly low HOM visibility.

In this work, we move beyond this conventional, source-only paradigm. 
By accounting for finite-time detection~\cite{Sasaki:06,Huang:10, Samara:21} without spectral filtering, we discover a \textit{detector-conditioned} heralded state whose purity can substantially exceed the JSA-derived limit, while maintaining high intrinsic heralding efficiency.

\begin{figure*}[t!]
\centering
\includegraphics[width=0.95\linewidth]{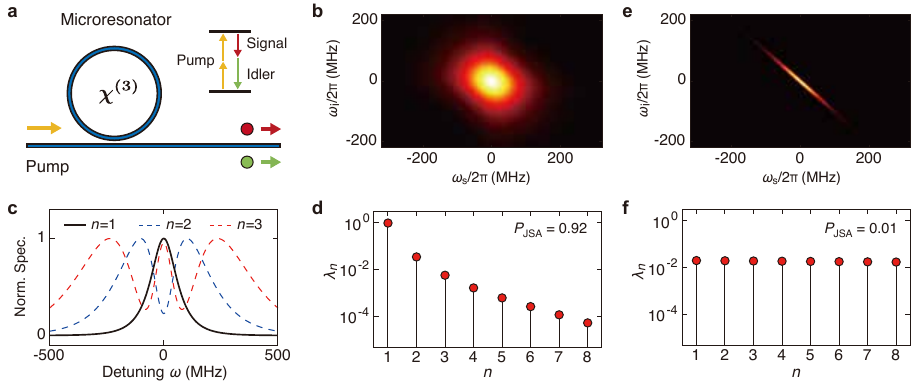}
\caption{
\textbf{Spectral purity of a microresonator-based photon-pair source.}
\textbf{a}, 
Schematic of photon-pair generation via cavity-enhanced SFWM in an integrated microresonator.
\textbf{b}, 
Simulated joint spectrum for a microresonator driven by a pulsed pump with 1-GHz bandwidth.
\textbf{c}, 
Normalized power spectra of the first three Schmidt modes ($n=1$--3) under pulsed pumping.
\textbf{d}, 
Schmidt coefficients $\lambda_n$ for modes $n=1$--8 under pulsed pumping, yielding a JSA-derived spectral purity of $P_\text{JSA}=0.92$.
\textbf{e}, 
Simulated, highly correlated, joint spectrum for a microresonator under CW pumping with 50-kHz linewidth.
\textbf{f}, 
Uniform distribution of Schmidt coefficients $\lambda_n$ under CW pumping, resulting in a vanishingly low spectral purity of $P_\text{JSA}=0.01$.
}
\label{Fig:1}
\end{figure*}

\textit{Photon spectral purity}---A photon-pair generated via a parametric nonlinear process is described by the state
\begin{equation}
|\Psi\rangle = \int \mathrm{d}\omega_s \mathrm{d}\omega_i\, F(\omega_s,\omega_i) a_s^\dagger(\omega_s) a_i^\dagger(\omega_i) |0\rangle,
\label{eqn:state}
\end{equation}
where $F(\omega_s,\omega_i)$ denotes the JSA, and $a_{s/i}^\dagger(\omega_{s/i})$ creates a signal or idler photon at frequency $\omega_{s/i}/2\pi$. 
Utilizing Schmidt decomposition \cite{Law:00}, the JSA is expressed as
\begin{equation}
F(\omega_s, \omega_i)=\sum_{n\in N^+} \sqrt{\lambda_n}\,\psi_n(\omega_s)\phi_n(\omega_i),
\label{eqn:decomposition}
\end{equation}
where $\lambda_n$ are the Schmidt coefficients, $\{\psi_n\}$ and $\{\phi_n\}$ are the two sets of orthonormal basis associated with the signal and idler modes.
By combining Eqs.~\eqref{eqn:state} and \eqref{eqn:decomposition}, and heralding on the idler photon, the resulting signal state $\rho_s = \sum_n \lambda_n |n_s\rangle\langle n_s|$ is mixed, 
where the signal modes are defined as $|n_s\rangle=\int\mathrm{d}\omega_s\, \psi_n(\omega_s)a^\dagger_s(\omega_s)|\mathrm{0}\rangle$.
The \textit{JSA-derived spectral purity} is
\begin{equation}
P_{\rm JSA}=\mathrm{Tr}\rho_s^2=\sum_{n\in N^+} \lambda_n^2.
\end{equation}
For a factorable JSA, the Schmidt coefficients reduce to a single non-zero value $\lambda_1=1$, yielding $P_{\rm JSA}=1$. 
For two identical, independent sources, $P_{\rm JSA}$ prescribes the maximum attainable HOM visibility. 
This framework has motivated JSA engineering for both bulk and integrated photon-pair sources \cite{Vernon:17, Zhong:18, Zhong:20, Paesani:20, Labonte:24, Alexander:25, Wang:26}. 

Figure~\ref{Fig:1}a illustrates photon-pair generation through cavity-enhanced spontaneous four-wave mixing (SFWM) \cite{Helt:10}. 
Two pump photons of frequency $\omega_p/2\pi$ are converted into a signal--idler pair constrained by energy conservation $2\omega_p=\omega_s+\omega_i$, where $\omega_{s/i}/2\pi$ represents the resonant frequency of the respective mode. 
For the remainder of this discussion, we refer $\omega_{p/s/i}/2\pi$ to the frequency detuning to the corresponding resonance. 
The generated state is described by the JSA:
\begin{equation}
F(\omega_s,\omega_i) \propto \,F_p(\omega_s+\omega_i)\,l(\omega_i)l(\omega_s),
\label{eqn:JSA}
\end{equation}
where $l(\omega_{s/i})=1/(\kappa/2-i\omega_{s/i})$ represents the Lorentzian resonance profile, 
$\kappa/2\pi$ is the resonance's full width at half maximum (FWHM).
The pump term is defined as
\begin{equation}
F_p(\omega) = \int \mathrm{d}\omega_p\,\alpha(\omega_p)\alpha(\omega-\omega_p)\,l(\omega_p)l(\omega-\omega_p),
\end{equation}
with $\alpha(\omega)$ being the spectral amplitude of the pump \cite{Vernon:17}.

Figure~\ref{Fig:1}b depicts the joint spectrum, $|F(\omega_s,\omega_i)|^2$, for a microresonator with $\kappa/2\pi=150$ MHz, driven by a pulsed pump with 1-GHz bandwidth. 
The corresponding power spectra for the first three Schmidt modes, i.e. $|\psi_n(\omega_s)|^2$ ($n=1$--3), are displayed in Fig.~\ref{Fig:1}c. 
The Schmidt coefficients $\lambda_n$ for modes $n=1$--8 are shown in Fig.~\ref{Fig:1}d, resulting in a spectral purity of $P_\text{JSA}=0.92$. 
Generally, Eq.~\eqref{eqn:JSA} predicts  $P_\text{JSA}<0.93$ under pulsed pumping where $\kappa/2\pi$ of the pump, signal, and idler resonances are identical \cite{Vernon:17}.
While strategies such as tunable coupling regions \cite{Vernon:17} or cascaded microresonator architectures \cite{Alexander:25} can improve $P_\text{JSA}$, they typically necessitate multiple active heaters to tune a single source, thereby increasing system complexity and introducing thermal crosstalk. 
Furthermore, achieving a factorable JSA often relies on pulsed pumping, which complicates deployment and increases costs.

In contrast, CW pumping eliminates the requirement for mode-locked lasers but produces highly correlated JSAs.
For a CW pump with 50-kHz linewidth, Figs.~\ref{Fig:1}e, f illustrate nearly uniform Schmidt coefficients, yielding a vanishingly low $P_{\rm JSA}=0.01$.
Within the source-only perspective, achieving high HOM visibility from such sources necessitates extreme narrowband filtering, incurring a prohibitive penalty in heralding efficiency.

\textit{Detector-conditioned purity}---The spectral purity $P_\text{JSA}$ characterizes the heralded photons without considering the influence of the detection process.
We therefore frame a model accounting for finite-time detection. 
The finite-time-detection scheme was proposed in Refs.~\cite{Sasaki:06,Huang:10} and demonstrated with microresonators in Ref.~\cite{Samara:21}.
Here we apply this method to cavity-enhanced SFWM, and investigate the heralded state.
To explore the scenario without spectral filtering, we assume a flat spectral acceptance window of bandwidth $B\gg\kappa$.
An idler detection event clicks within an idler-photon detection window of width $T$, defining a band-limited, finite-time detection operator
\begin{equation}
\hat{M}_T = \sum_{n\in N^+} \mu_n |d_n\rangle\langle d_n|.
\end{equation}
Here, $\mu_n$ represents the acceptance weight of the $n$-th detection mode within the interval $[-T/2,T/2]$, 
and $\{|d_n\rangle=\int\mathrm{d}\omega\chi_n(\omega)a^\dagger(\omega)|0\rangle\}$ forms an orthonormal basis of the detection modes.
The eigenvalues $\mu_n$ and eigenmodes $\chi_n(\omega)$ are determined by
\begin{equation}
    \int\mathrm{d}\omega' M_T(\omega, \omega') \chi_n(\omega') = \mu_n\chi_n(\omega),
\end{equation}
where the detection kernel is
\begin{equation}
M_T(\omega, \omega') = \frac{T}{2\pi} \mathrm{sinc}\frac{T(\omega-\omega')}{2\pi},
\label{eqn:MT}
\end{equation}
with $\omega, \omega'\in [-B/2, B/2]$, 
and $\mathrm{sinc}(x)=\sin(\pi x)/(\pi x)$. 
Given $B\gg\kappa$, the spectral bandwidth $B$ does not impose purity-enhancing filtering. 
More details are provided in \cite{SuppMat} Note 1A.

Conditioning the biphoton state $|\Psi\rangle$ on an idler detection event yields the unnormalized signal state
\begin{equation}
\tilde{\rho}_s
=\mathrm{Tr}_i\!\left[\hat{M}_{T}^{(i)}\,|\Psi\rangle\langle\Psi|\right].
\label{eqn:rho_heralded}
\end{equation}
The normalized heralded signal state is then $\rho_s=\tilde{\rho}_s/\mathrm{Tr}\,\tilde{\rho}_s$,
where the superscript $(i/s)$ denotes the operator acting on the idler or signal photon.
The \textit{detector-conditioned purity} of the heralded signal photon is
\begin{equation}
P=\mathrm{Tr}\,\rho_s^2, 
\label{eqn:rho_purity}
\end{equation}
The corresponding intrinsic heralding efficiency is
\begin{equation}
\eta=\mathrm{Tr}\!\left[\hat{M}_{T'}^{(s)}\,\rho_s\right],
\end{equation}
where $T'$ is the signal-photon detection-window width. 
For heralded photons from two independent and identical sources, $P(T)$ establishes the upper bound of the experimentally accessible HOM visibility $V$. 
Here $\eta$, determined by $T'$, represents the probability of detecting the signal upon an idler click.

\begin{figure}[t!]
\centering
\includegraphics[width=\columnwidth]{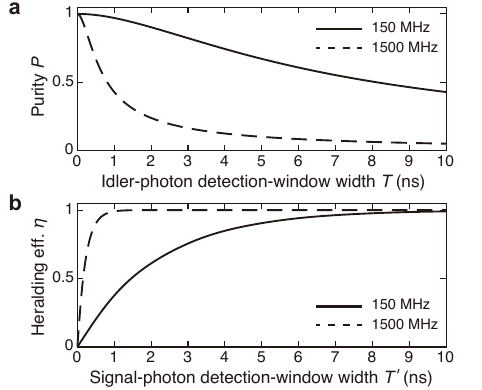}
\caption{
\textbf{Detector-conditioned purity and intrinsic heralding efficiency.}
\textbf{a}, 
Detector-conditioned purity $P$ as a function of $T$ for $\kappa/2\pi=150$ and $1500$ MHz.
\textbf{b}, 
Intrinsic heralding efficiency $\eta$ as a function of $T'$ for $\kappa/2\pi=150$ and $1500$ MHz, with $T=100$ ps.
}
\label{Fig:2}
\end{figure}

\begin{figure*}[t!]
\centering
\includegraphics[width=0.95\linewidth]{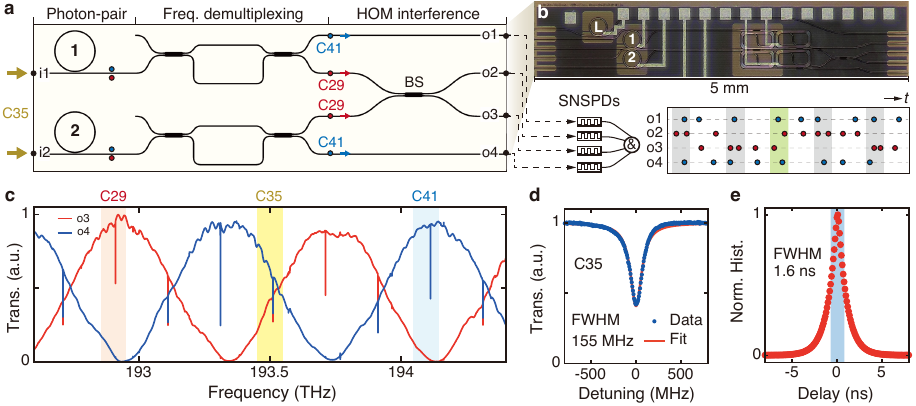}
\caption{
\textbf{Chip layout and experimental characterization.}
\textbf{a}, 
Schematic of the chip layout and detection scheme.
Two microresonators are pumped at the C35 resonance (193.5 THz), generating C29 signal and C41 idler photon pairs.
UMZIs demultiplex these channels, routing C41 idlers to ports o1 and o4 to herald the two C29 signal photons, which interfere on the BS. 
All four outputs are detected by SNSPDs for coincidence analysis.
Inset: standard single-window coincidence logic, where green windows represent registered fourfold events and gray windows indicate HOM-suppressed cases.  
\textbf{b}, 
Optical micrograph of the Si$_3$N$_4$ photonic integrated circuits, showing two photon-source microresonators and an auxiliary microresonator (L) for PDH stabilization of the CW pump.
\textbf{c}, 
Normalized transmission spectra at the UMZI outputs, demonstrating complementary routing of the C29 and C41 channels.
\textbf{d}, 
The C35 resonance profile for microresonator~2, with a Lorentzian fit yielding a FWHM of $\kappa/2\pi=155$ MHz.
\textbf{e}, 
Normalized signal--idler coincidence histogram exhibiting a temporal FWHM of 1.6 ns. 
}
\label{Fig:3}
\end{figure*}

Using Eqs.~\eqref{eqn:rho_heralded} and \eqref{eqn:rho_purity}, we calculate $P(T)$ for a microresonator source under a 50-kHz-linewidth CW pump, as shown in Fig.~\ref{Fig:2}a. 
Reducing $T$ suppresses higher-order detection modes, thereby increasing $P$.
For instance, a microresonator of $\kappa/2\pi=1500$ MHz requires $T=56$~ps to reach $P=0.99$, while a microresonator with $\kappa/2\pi=150$ MHz achieves the same $P=0.99$ at increased $T=0.56$ ns. 
The latter $T=0.56$ ns is above the typical time resolution of common detectors (e.g. $\sim0.1$ ns for our detectors, described later), validating our detector-conditioned method. 
Consequently, high-$Q$ microresonators significantly relax the stringent requirement of ultralow-jitter detectors.  
Simulation details are provided in \cite{SuppMat} Note 1B.

Figure~\ref{Fig:2}b further shows $\eta$ as a function of $T'$, corresponding to the fraction of the heralded signal-photon wave-packet captured within the signal window. 
For $\kappa/2\pi=150$ MHz (1500 MHz), $\eta$ exceeds 0.99 when $T'>4.9$ ns (0.49 ns).
Notably, Eq.~\eqref{eqn:rho_heralded} shows that $P$ is independent of $T'$. 
Therefore, decreasing $T$ leads to $P\rightarrow1$; independently, increasing $T'$ results in $\eta\rightarrow1$.


\textit{Experimental demonstration}---We demonstrate efficient heralded-photon generation and verify their near-unity $P$ through chip-scale HOM interference between two independent CW-driven microresonator sources.
The integrated circuit is fabricated on an ultralow-loss silicon nitride (Si$_3$N$_4$) platform \cite{Liu:21, Ye:23}, enabling both high-$Q$ microresonators and scalable quantum circuits \cite{Alexander:25, Aghaee:25, Chen:26}.
As illustrated in Fig.~\ref{Fig:3}a, the chip integrates two microresonators, two unbalanced Mach-Zehnder interferometers (UMZIs) for frequency demultiplexing, and a multimode interferometer (MMI) \cite{Soldano:95} serving as the beam splitter (BS) for HOM interference. 
An optical micrograph of the 5-mm-wide chip is shown in Fig.~\ref{Fig:3}b, containing integrated heaters for independent thermal tuning of each microresonator and UMZI. 
Fabrication details are provided in \cite{SuppMat} Note 2.

Both photon-source microresonators are driven by a single CW laser tuned to the ITU C35 channel (193.5 THz). 
Signal and idler photons are collected from resonances in the C29 (192.9 THz) and C41 (194.1 THz) channels, respectively.
Each source is pumped with an on-chip power of 0.5~mW. 
An auxiliary microresonator, marked as ``L'' in Fig.~\ref{Fig:3}b, provides Pound--Drever--Hall (PDH) stabilization of the pump laser \cite{Drever:83}. 
Following each source, a UMZI demultiplexes the signal and idler photons. 
We characterize the demultiplexing from port i2 to ports o3 and o4 using a vector spectrum analyzer \cite{Luo:24}.
As shown in Fig.~\ref{Fig:3}c, the two outputs exhibit complementary transmission at C29 and C41.
The C41 idler photons are routed off-chip to serve as heralds, while the heralded C29 signal photons interfere at the on-chip BS. 
All four outputs are directed to superconducting nanowire single-photon detectors (SNSPDs).
Photon-pair generation is verified by the coincidence histogram in Fig.~\ref{Fig:3}e, showing a correlation peak with a FWHM of 1.6~ns. 
This FWHM reflects the photon ringdown time and is determined by the resonance linewidth $\kappa/2\pi$. 
A Lorentzian fit to a representative resonance (C35 of microresonator 2) yields $\kappa/2\pi=155$ MHz. 
The corresponding C29 and C41 resonances exhibit comparable $\kappa/2\pi$, consistent with the observed 1.6-ns coincidence FWHM. 
More setup and device details are provided in \cite{SuppMat} Notes 3 and 4.

\begin{figure*}[t!]
\centering
\includegraphics[width=0.95\linewidth]{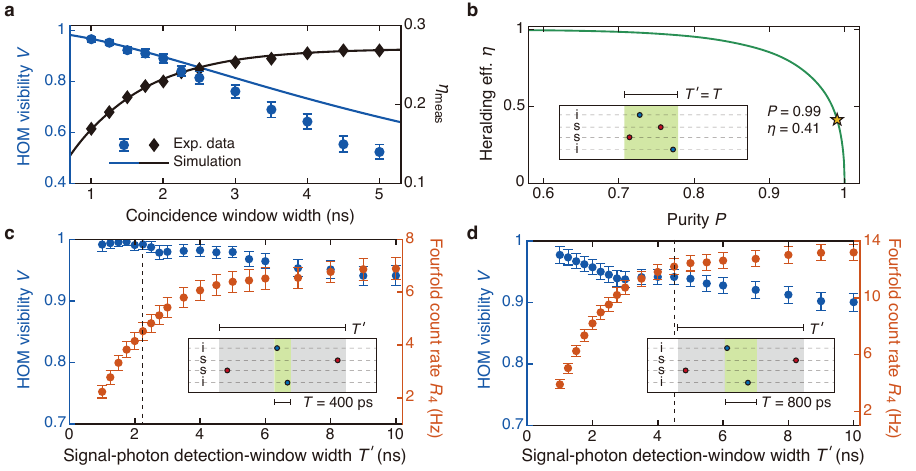}
\caption{
\textbf{Photon-pair performance.}
\textbf{a}, 
Measured $V$ and $\eta_\mathrm{meas}$ versus the coincidence window width under single-window coincidence logic with $\kappa/2\pi=155$~MHz. 
\textbf{b}, 
Purity--efficiency trade-off under single-window coincidence logic.
Inset, fourfold coincidence under single-window coincidence logic used in \textbf{a} and \textbf{b}, with signals and idlers sharing identical detection-window width. 
\textbf{c, d}, 
Measured $V$ and $R_4$ versus $T'$ for $T=400$ and 800~ps, respectively. 
Insets, fourfold coincidence under double-window coincidence logic, enabling independently configured $T$ and $T'$. 
Dashed lines mark the operating points $V=0.992(8)$ at $R_4=4.5(3)$~Hz and $V=0.942(12)$ at $R_4=12.2(6)$~Hz.
All reported $V$ values are obtained without background or accidental count subtraction.
The error bars represent one standard deviation, calculated from Poissonian counting statistics of the raw detection events.
}
\label{Fig:4}
\end{figure*}

Next, we verify that the measured HOM visibility $V$ can reach the theoretically predicted $P$. 
Experimentally, $V$ is extracted from fourfold coincidences recorded using a configurable coincidence window whose width defines $T$.
As illustrated in Fig.~\ref{Fig:3}a inset, a fourfold event is registered when both pairs (four photons) are detected within the same window, marked in green.
HOM interference causes the two signal photons to exit through the same output channel, thereby eliminating the corresponding fourfold coincidences, marked in gray.
We obtain $V$ by comparing the zero-delay fourfold counts with a baseline measured at 100-ns delay, followed by correction for BS imbalance.
All reported $V$ values are obtained \emph{without} background or accidental count subtraction.

Figure~\ref{Fig:4}a blue symbols present $V$ as a function of coincidence window width. 
For the window width below 2~ns, the $V$ data closely follow the upper bound set by $P$ with $\kappa/2\pi=155$ MHz (blue curve). 
Once exceeding the 1.6-ns coincidence FWHM, the window starts to accept background noises that are not included in the simulation, degrading $V$.

The coincidence window also determines $T'$, detailed in \cite{SuppMat} Note 5.
This constraint results in a purity--efficiency trade-off shown in Fig.~\ref{Fig:4}b, where $P>0.99$ bounds $\eta<41\%$. 
The measured heralding efficiency $\eta_{\rm meas}$, shown by Fig.~\ref{Fig:4}a black symbols, agrees with the predicted $L\cdot\eta$ (black curve), where $L=0.27$ is the external link efficiency, detailed in \cite{SuppMat} Note 3. 

This purity--efficiency trade-off limits the utility of CW-driven microresonators for deterministic heralding of single photons or EPR states.
However, we find that this trade-off is not intrinsic to the source, but arises from the \textit{single-window} coincidence logic, as illustrated in Fig. \ref{Fig:4}b inset with $T=T'$ for fourfold coincidence.
This trade-off can be overcome by independently configuring $T$ and $T'$, as implemented by the \textit{double-window} coincidence logic illustrated in the insets of Figs.~\ref{Fig:4}c and \ref{Fig:4}d, detailed in \cite{SuppMat} Note 5.
The measured $V$ and delayed-baseline fourfold count rates $R_4$ as functions of $T'$ are shown in Figs.~\ref{Fig:4}c and \ref{Fig:4}d for $T=400$ and 800~ps, respectively. 
As $T'$ increases, $V$ decreases marginally, in contrast to the trend in Fig.~\ref{Fig:4}a. 
Meanwhile, $R_4$ increases substantially, and then saturates when the full signal-photon wave-packets are captured. 
For $T=400$ and 800 ps, $V=0.992(8)$ and 0.942(12) are achieved at $R_4=4.5(3)$ and $12.2(6)$~Hz, marked by the dashed lines. 

In conclusion, we have demonstrated that CW-driven, high-$Q$ microresonators generate indistinguishable heralded photons without spectral filtering or source engineering, while preserving high heralding efficiency.
This capability relies on the photon-pair correlation time being sufficiently long to be resolved by the detector. 
The resulting HOM visibility $V$ and fourfold count rate $R_4$ are compared with previous works based on integrated photonics, detailed in \cite{SuppMat} Note 6.
By further leveraging CW pumping, this architecture circumvents the need for complex inter-source synchronization and remains inherently compatible with chip-scale lasers and heterogeneous integration \cite{Xiang:21}.
These attributes establish high-$Q$ microresonators as a robust and practical foundation for scalable photon sources in quantum networks and light--matter interfaces.

\vspace{3mm}
\begin{acknowledgments}
\emph{Acknowledgments}---We thank Zhenyuan Shang, Jiahao Sun and Chen Shen for assisting Si$_3$N$_4$ chip fabrication. 
We acknowledge support from the National Key R\&D Program of China (Grant No. 2024YFA1409300), 
Quantum Science and Technology–National Science and Technology Major Project (Grant No. 2023ZD0301500), 
National Natural Science Foundation of China (Grant No. 12404417 and U25D9005), 
Shenzhen Science and Technology Program (Grant No. RCJC20231211090042078), 
and Shenzhen-Hong Kong Cooperation Zone for Technology and Innovation (HZQBKCZYB2020050). 
Y.-H. L. acknowledges support from the Young Elite Scientists Sponsorship Program by CAST (Grant No. YESS20240475). 
Y.-H. L. conceived the experiment. 
R. C. and Y.-H. L. designed the chips and performed the theoretical calculations.
S. H., Z. Z., Z. C. and X. B. fabricated the chips.
R. C., Y.-H. L., S. Z. and Y. C. built the setup and performed the experiment.
R. C., Y.-H. L. and J. L. analyzed the data and wrote the manuscript. 
J. L. supervised the project.

\emph{Data availability}---The code and data used to produce the plots within this work will be released on the repository Zenodo upon publication of this manuscript.
\end{acknowledgments}

\bibliographystyle{apsrev4-1}
\bibliography{bibliography, bib_add, Supp}

\end{document}


\title{Supplementary Material for: Filter-Free Indistinguishable Photon Generation from\\Continuous-Wave-Driven Integrated Microresonators}

\author{Ruiyang Chen}
\affiliation{School of Physical Sciences and Hefei National Laboratory, University of Science and Technology of China, Hefei 230026, China}
\affiliation{International Quantum Academy and Shenzhen Futian SUSTech Institute for Quantum Technology and Engineering, Shenzhen 518048, China}

\author{Sicheng Zeng}
\affiliation{International Quantum Academy and Shenzhen Futian SUSTech Institute for Quantum Technology and Engineering, Shenzhen 518048, China}
\affiliation{Shenzhen Institute for Quantum Science and Engineering, Southern University of Science and Technology, Shenzhen 518055, China}

\author{Yuan Chen}
\affiliation{International Quantum Academy and Shenzhen Futian SUSTech Institute for Quantum Technology and Engineering, Shenzhen 518048, China}

\author{Sanli Huang}
\affiliation{School of Physical Sciences and Hefei National Laboratory, University of Science and Technology of China, Hefei 230026, China}
\affiliation{International Quantum Academy and Shenzhen Futian SUSTech Institute for Quantum Technology and Engineering, Shenzhen 518048, China}

\author{Zeying Zhong}
\affiliation{International Quantum Academy and Shenzhen Futian SUSTech Institute for Quantum Technology and Engineering, Shenzhen 518048, China}
\affiliation{Shenzhen Institute for Quantum Science and Engineering, Southern University of Science and Technology, Shenzhen 518055, China}

\author{Zhen Chen}
\affiliation{School of Physical Sciences and Hefei National Laboratory, University of Science and Technology of China, Hefei 230026, China}
\affiliation{International Quantum Academy and Shenzhen Futian SUSTech Institute for Quantum Technology and Engineering, Shenzhen 518048, China}

\author{Xue Bai}
\affiliation{International Quantum Academy and Shenzhen Futian SUSTech Institute for Quantum Technology and Engineering, Shenzhen 518048, China}
\affiliation{Qaleido Photonics, Shenzhen 518048, China}

\author{Yi-Han Luo}
\email{luoyh@iqasz.cn}
\affiliation{International Quantum Academy and Shenzhen Futian SUSTech Institute for Quantum Technology and Engineering, Shenzhen 518048, China}

\author{Junqiu Liu}
\affiliation{School of Physical Sciences and Hefei National Laboratory, University of Science and Technology of China, Hefei 230026, China}
\affiliation{International Quantum Academy and Shenzhen Futian SUSTech Institute for Quantum Technology and Engineering, Shenzhen 518048, China}

\maketitle
\tableofcontents
\clearpage

\section{Details of the theoretical model}
\vspace{0.2cm}

\subsection{Detection kernel}

The detection process can be modeled as a projection of the incident photon onto a single-photon state.
The state is jointly determined by the detector's spectral acceptance and detection time window.
For an ideal rectangular spectral acceptance window of bandwidth $B$, the state can be written as
\begin{equation}
|d\rangle = \int\mathrm{d}\omega\, \chi(\omega)|\omega\rangle,
\label{eqn:det-state}
\end{equation}
where $\chi(\omega)$ is the spectral amplitude and vanishes outside the spectral acceptance window $\omega\in[-B/2,B/2]$.
The corresponding wavefunction in the time domain is
\begin{equation}
\psi(t) = \frac{1}{\sqrt{2\pi}}\int \mathrm{d}\omega\, \chi(\omega)e^{-i\omega t}.
\label{eqn:det-state-time}
\end{equation}

We further consider the finite-time detection, namely, the detector is active only within a detection window of width $T$ and is insensitive to photons arriving outside this window.
The detector clicking probability is characterized by $\int_{-T/2}^{T/2}\mathrm{d}t\, |\psi(t)|^2$. 
Substituting Eq.~\eqref{eqn:det-state-time} into this expression, one obtains
\begin{equation}
    \int_{-T/2}^{T/2}\mathrm{d}t\, |\psi(t)|^2 = \int\mathrm{d}\omega\mathrm{d}\omega'\, \chi(\omega) M_T(\omega, \omega') \chi^*(\omega')
\end{equation}
where $M_T(\omega, \omega') = \frac{1}{2\pi}\int_{-T/2}^{T/2}\mathrm{d}t\, e^{-i(\omega-\omega')t}$ is the detection kernel associated with the detection window.
Evaluating the time integral gives
\begin{equation}
    M_T(\omega, \omega') = \frac{\sin\left[(\omega-\omega')T/2\right]}{\pi(\omega-\omega')}. 
\end{equation}
Equivalently, the kernel can be written in sinc form as
\begin{equation}
M_T(\omega,\omega')=\frac{T}{2\pi}\mathrm{sinc}\left[\frac{T(\omega-\omega')}{2\pi}\right],
\end{equation}
defined on the domain $\omega,\omega'\in[-B/2,B/2]$. 
Here, the sinc function is defined as $\mathrm{sinc}(x)=\sin(\pi x)/(\pi x)$.

In experiment, we are interested in the probability that an incident photon in the mode $\chi(\omega)$ triggers the detector.
This probability is given by the functional
\begin{equation}
    \mu[\chi] = \frac{\int\mathrm{d}\omega\mathrm{d}\omega'\, \chi(\omega) M_T(\omega, \omega') \chi^*(\omega')}{\int\mathrm{d}\omega\, |\chi(\omega)|^2},
\end{equation}
Equivalently, this functional can be written in operator form as
\begin{equation}
    \mu = \frac{\langle d|\hat{M}_T| d\rangle}{\langle d|d\rangle}
\end{equation}
where $|d\rangle$ is the single-photon state defined in Eq.~\eqref{eqn:det-state}, and
\begin{equation}
\hat{M}_T=\int \mathrm{d}\omega\mathrm{d}\omega' M_T(\omega, \omega')|\omega\rangle\langle\omega'|
\end{equation}
is the Hermitian operator associated with the kernel $M_T(\omega,\omega')$.

The functional $\mu[\chi]$ is a Rayleigh quotient of the kernel $M_T(\omega,\omega')$, thus
its extremal values are obtained when $\chi_n(\omega)$ is an eigenfunction of $M_T(\omega,\omega')$, satisfying
\begin{equation}
\int\mathrm{d}\omega'\,M_T(\omega,\omega')\,\chi_n(\omega')=\mu_n\,\chi_n(\omega).
\end{equation}
The eigenfunctions ${\chi_n}$ form an orthogonal set, and the corresponding eigenvalues $\mu_n$ quantify the detector click probability for the associated detection modes.
Notably, the measurement kernel can be decomposed as
\begin{equation}
M_T(\omega,\omega')=\sum_n \mu_n\, \chi_n(\omega)\chi_n^*(\omega').
\label{eqn:kernel-decom}
\end{equation}
Correspondingly, the operator $\hat{M}_T$ can be written in the form
\begin{equation}
\hat{M}_T=\sum_n \mu_n\, |d_n\rangle\langle d_n|,
\label{eqn:det-measure}
\end{equation}
where 
\begin{equation}
|d_n\rangle=\int \mathrm{d}\omega\, \chi_n(\omega)|\omega\rangle
\end{equation}
is the $n$th detection eigenstate.

\subsection{Detector-conditioned purity and intrinsic heralding efficiency simulation}

We now introduce the calculation of the detector-conditioned state purity $P$ and intrinsic heralding efficiency $\eta$.
The biphoton state generated via a parametric nonlinear process can be written as
\begin{equation}
    |\Psi\rangle = \int\mathrm{d}\omega_s\mathrm{d}\omega_i\, F(\omega_s, \omega_i)\,|\omega_s, \omega_i\rangle, 
\end{equation}
where $F(\omega_s,\omega_i)$ is the joint spectral amplitude (JSA) of the photon pair.
For the cavity-enhanced spontaneous four-wave mixing considered here, the explicit form of $F(\omega_s,\omega_i)$ is given in the main text.

A click on the idler heralds the presence of the signal.
Using the idler-side measurement operator $\hat{M}_T^{(i)}$ defined as Eq. \ref{eqn:det-measure}, the unnormalized heralded state of the signal photon is obtained as
\begin{equation}
    \tilde{\rho}_s = \mathrm{Tr}_i \left[ \hat{M}_T^{(i)} |\Psi\rangle\langle\Psi|\right] = \sum_n \mu_n |\varphi_n\rangle \langle\varphi_n|
\end{equation}
where $|\varphi_n\rangle = {}_i\langle d_n|\Psi\rangle$. 
Substituting the expression for $|d_n\rangle$ into the above equation gives
\begin{equation}
    |\varphi_n\rangle = \int \mathrm{d}\omega_s\mathrm{d}\omega \, F(\omega_s, \omega) \chi_n^*(\omega) |\omega_s\rangle.
\end{equation}
Therefore, the unnormalized heralded state can be written explicitly as
\begin{equation}
\tilde{\rho}_s = \sum_n \mu_n \int\,\mathrm{d}\omega_s\mathrm{d}\omega_s'\mathrm{d}\omega\mathrm{d}\omega' F(\omega_s, \omega)\chi_n^*(\omega)\chi_n(\omega')F^*(\omega_s', \omega')|\omega_s\rangle\langle\omega_s'|
\end{equation}
Using the kernel decomposition in Eq.~\eqref{eqn:kernel-decom}, we obtain the following integral form of the unnormalized heralded signal state
\begin{equation}
\tilde{\rho}_s = \int\,\mathrm{d}\omega_s\mathrm{d}\omega_s'\mathrm{d}\omega\mathrm{d}\omega' F(\omega_s, \omega)M_T(\omega, \omega')F^*(\omega_s', \omega')|\omega_s\rangle\langle\omega_s'|,
\end{equation}
which can be directly discretized for numerical calculation.
The normalized heralded state is given by $\rho_s=\tilde{\rho}_s/\mathrm{Tr}(\tilde{\rho}_s)$.
Subsequently, the corresponding detector-conditioned purity of the signal photon is
\begin{equation}
P=\mathrm{Tr}\rho_s^2, 
\end{equation}
which characterizes the indistinguishability of the heralded signal photon and determines the accessible Hong--Ou--Mandel (HOM) visibility between two identical and independent heralded single photons.

Another important figure of merit is the intrinsic heralding efficiency $\eta$, defined as the probability of detecting the signal photon upon an idler click.
As the heralded signal photon is described by $\rho_s$, $\eta$ is the probability with $\rho_s$ being detected by $\hat{M}_{T'}^{(s)}$, namely,
\begin{equation}
    \eta = \mathrm{Tr} \left(\hat{M}_{T'}^{(s)}\rho_s\right).
\end{equation}
Generally, $\hat{M}_{T'}^{(s)}$ is associated with signal detection window $T'$, and can be configured independently with $\hat{M}_{T}^{(i)}$.

\clearpage

\section{Device fabrication process}
\vspace{0.2cm}

\begin{figure*}[h!]
\centering
\includegraphics[width=0.95\textwidth]{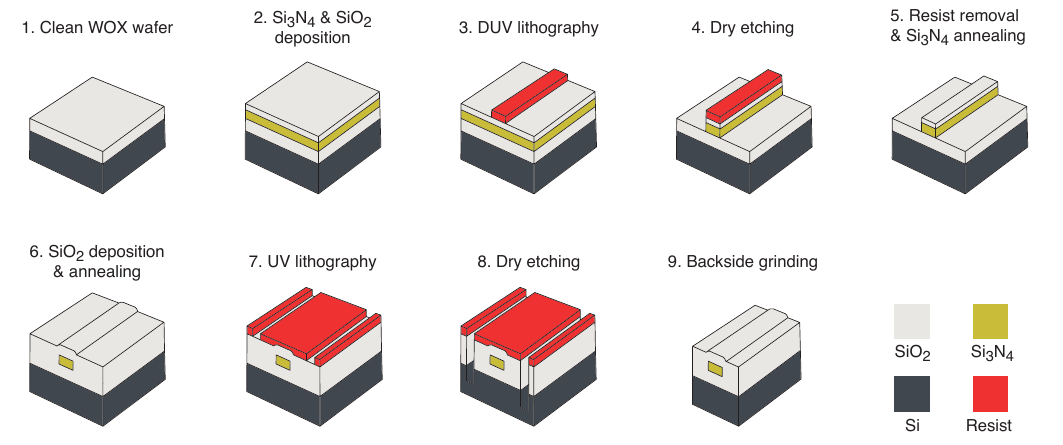}
\caption{
\textbf{The DUV subtractive process flow of 6-inch-wafer Si$_3$N$_4$ photonic integrated chip foundry fabrication.}
WOX, thermal wet oxide (SiO$_2$). }
\label{Fig:S0}
\end{figure*}

The Si$_3$N$_4$ photonic integrated chips (PICs) are fabricated using an optimized deep-ultraviolet (DUV) subtractive process on 6-inch wafers \cite{Ye:23, Sun:25}, performed in our CMOS foundry.
The process flow is illustrated in Fig. \ref{Fig:S0}. 
First, an 800-nm-thick Si$_3$N$_4$ film is deposited on a clean thermal wet SiO$_2$ substrate via low-pressure chemical vapor deposition (LPCVD).
Subsequently, an SiO$_2$ film is deposited on the Si$_3$N$_4$ as an etch hardmask, again via LPCVD.
Next, DUV stepper lithography is performed, followed by dry etching to transfer the pattern from the DUV photoresist to the SiO$_2$ hardmask, and then to the Si$_3$N$_4$ layer.
The dry etching uses etchants comprising CHF$_3$ and O$_2$ to create ultra-smooth and vertical etched sidewalls, critical for minimizing optical losses in waveguides.
Afterward, the photoresist is removed, and thermal annealing of the entire wafer is applied under a nitrogen atmosphere at 1200 $^\circ$C.
Then a 3-$\mu$m-thick SiO$_2$ top cladding layer is deposited on the wafer, followed by another thermal annealing at 1200 $^\circ$C.
Finally, UV photolithography and deep dry etching are performed to create smooth chip facets facilitating fiber coupling. 

After completion of the photonic-circuit fabrication process and before grinding or dicing, the wafer surface is planarized by chemical--mechanical polishing (CMP).
A TiN thin film is deposited and patterned to form resistive thermo-optic heaters above selected resonators and interferometers, following TiN-heater integration schemes reported in active Si$_3$N$_4$ photonic platforms \cite{Yong:22}.
The TiN heaters are embedded in an additional LPCVD SiO$_2$ layer.
Contact vias are opened through this oxide, and a top Al metallization layer is patterned to connect the heaters to wire-bond pads.
The wafer is separated into individual chips through backside grinding or dicing.

\section{Experimental setup}

\begin{figure*}[h!]
\centering
\includegraphics[width=\textwidth]{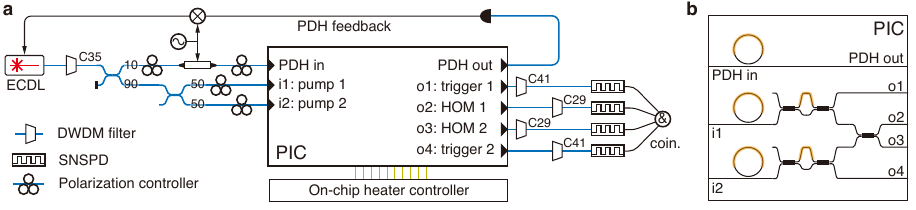}
\caption{
\textbf{Experimental setup.}
\textbf{a}, Fiber-to-chip setup. 
The ECDL output is transmitted through the C35 channel and subsequently split into a PDH-locking arm and a photon-pair-generation arm.
The PDH arm is coupled to a reference microresonator through the \emph{PDH in/out} ports to lock the laser frequency.
The photon-pair-generation arm is equally split and injected into two on-chip microresonators through \emph{pump 1} and \emph{pump 2}.
The generated idler photons are detected from \emph{trigger 1} and \emph{trigger 2} after C41 filtering, whereas the heralded signal photons interfere on an on-chip BS and are detected from \emph{HOM 1} and \emph{HOM 2} after C29 filtering.
The photons are detected by SNSPDs.
The detection events are analyzed with a time tagger. 
\textbf{b}, Definitions of the PIC ports and their corresponding connections to the on-chip routing.
}
\label{Fig:S-setup}
\end{figure*}

The experimental setup is shown in Fig.~\ref{Fig:S-setup}a, and the corresponding on-chip port connections are detailed in Fig.~\ref{Fig:S-setup}b.
This section focuses on the experimental apparatus and its interface with the PIC.
Detailed characterizations of the integrated devices are presented in the next section.

An external-cavity diode laser (ECDL, Toptica CTL 1550) is first filtered by a dense wavelength-division multiplexing (DWDM) filter centered at 193.5~THz, corresponding to the International Telecommunication Union (ITU) C35 channel.
This filter suppresses the laser sideband noise by approximately 100~dB.
After filtering, 10\% of the laser power is sent to the \emph{PDH in} port of the photonic integrated circuit (PIC).
This port is connected to a reference microresonator that is identical to the photon-pair-generation ones.
Light from the \emph{PDH out} port is used to generate a Pound--Drever--Hall (PDH) error signal, which is fed back to stabilize the laser frequency and lock it to a resonance of the reference microresonator.

The remaining 90\% of the laser power is equally split and injected into the \emph{pump 1} and \emph{pump 2} ports.
These two ports pump two source microresonators on the same chip and generate photon pairs through cavity-enhanced spontaneous four-wave mixing. 
The generated photon pairs are separated on chip by unbalanced Mach--Zehnder interferometers (UMZIs).
The idler photons are routed to the \emph{trigger 1} and \emph{trigger 2} ports.
After external DWDM filtering at C41, they are directly detected by superconducting nanowire single-photon detectors (SNSPDs) and serve as triggers for the two signal photons.
The heralded signal photons are routed to an on-chip multimode interference (MMI) as a beam splitter (BS), where they overlap and interfere.
The two outputs of the MMI are collected from ports \emph{HOM 1} and \emph{HOM 2}.
After external DWDM filtering at the C29 channel, the signal photons are detected by SNSPDs.
The C29/C41 output DWDM filters have 40-GHz passbands, much broader than the microresonator linewidths ($\approx 150$~MHz), satisfying $B\gg\kappa$. 
These filters serve only for channel selection and residual-pump rejection, not for spectral shaping or purity-enhancing filtering.

The multi-channel input and output coupling of the PIC is implemented using V-groove-mounted UHNA-4 fiber arrays with a pitch of 127~$\mu$m.
The UHNA-4 fibers are thermally expanded and spliced to standard SMF-28e fibers for connection to the subsequent fiber components.
With an index-matching liquid of refractive index 1.50, the average fiber-to-chip coupling efficiency reaches approximately 55\%.
The photon extraction efficiency from the microresonators is approximately 75\%; the on-chip circuit transmission is approximately 90\%; the DWDM filter transmission is approximately 85\%; and the SNSPD detection efficiency is approximately 85\%.
Combining these contributions gives an overall link efficiency of approximately $55\%\times75\%\times90\%\times85\%\times85\%=27\%$ ($-5.5$ dB overall insertion loss).

The three microresonators and two UMZIs shown in Fig.~\ref{Fig:S-setup}b are thermally tuned by integrated heaters.
Each heater is driven by an independent voltage source with a 0--20~V tuning range and a 16-bit voltage resolution.
The microresonator heaters are used to align the resonance frequencies of the signal modes involved in the HOM interference; the UMZI heaters tune the 800-GHz-FSR UMZIs to demultiplex the C29 and C41 channels. 
The PIC temperature is stabilized to approximately 1~mK.
Once the laser is locked to the reference microresonator, the resonances of the source microresonators can be aligned to the laser frequency solely by tuning their on-chip heaters, without any additional active feedback.

\section{Characterization of the microresonators and UMZIs}

We characterize the two microresonator--UMZI cascades with a vector spectrum analyzer \cite{Luo:24}.
Each cascade consists of a microresonator, featuring a 200-GHz free spectral range (FSR), used for photon-pair generation followed by a UMZI used for signal--idler demultiplexing.
For the upper cascade, light is launched into i1 and collected from o1 and o2; for the lower cascade, light is launched into i2 and collected from o3 and o4, following the port definition in Fig.~\ref{Fig:S-setup}b.

\begin{figure}[h!]
    \centering
    \includegraphics[width=0.98\linewidth]{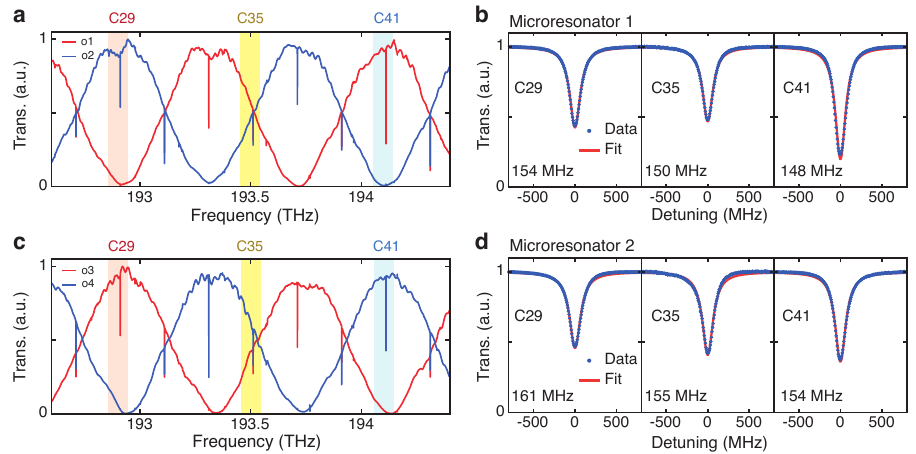}
    \caption{
    \textbf{Characterization of microresonators and UMZIs.}
    \textbf{a,} Transmission spectra measured from input port i1 to output ports o1 and o2.
    \textbf{c,} Transmission spectra measured from input port i2 to output ports o3 and o4.
    \textbf{b,d,} Lorentzian fits to the resonances in C29, C35, and C41 extracted from panels \textbf{a} and \textbf{c}, respectively.
    The fitted resonance FWHMs are labeled in each panel.
    }
    \label{Fig:S5}
\end{figure}

The measured transmission spectra contain the microresonator resonances modulated by the broad UMZI transfer envelope, as shown in Fig.~\ref{Fig:S5}a,c.
The envelopes reveal a UMZI FSR of 800~GHz.
The signal and idler channels used in the experiment, C29 and C41, are separated by 1200~GHz, corresponding to 1.5 UMZI FSRs.
They are therefore routed to complementary output ports, confirming the on-chip signal--idler demultiplexing used in the HOM measurement.

We then extract the resonances from the transmission spectra.
Figures~\ref{Fig:S5}b,d show the resonances fitted with Lorentzian functions for microresonators 1 and 2, respectively.
The loaded linewidths are 148--161~MHz, consistent with the value used in the detector-conditioned temporal model.
These linewidths lead to nanosecond-scale photon coherence times, enabling the finite-time detection---previously demonstrated in atomic systems\cite{Peng:16}---to be practically implemented with integrated photonics.

\section{Coincidence logic}

\begin{figure*}[h!]
\centering
\includegraphics[width=0.85\textwidth]{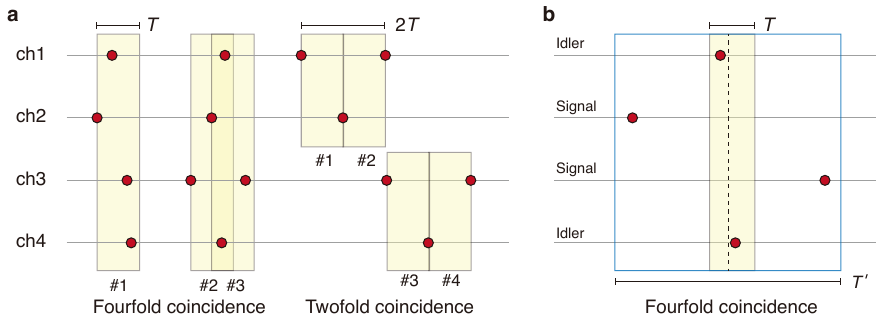}
\caption{
\textbf{Coincidence logic.}
\textbf{a}, Under standard single-window logic, fourfold coincidence requires all four clicks to lie within the same coincidence window, whose width determines the idler and signal detection window $T$ and $T'=T$.
For twofold coincidence, conditioning on idler yields an effective acceptance interval \([-T,T]\) for the signal, corresponding to \(T'=2T\).
\textbf{b}, Under double-window logic, the two idler clicks are constrained within \(T\), while the two signal clicks are accepted within an independently configured window \(T'\) centered at the mean idler-click time.
}
\label{Fig:S-coincidence}
\end{figure*}

In the experiment, coincidence events are identified by a time tagger (Swabian Time Tagger Ultra), whose \emph{standard} single-window coincidence logic is illustrated in Fig.~\ref{Fig:S-coincidence}a.
For an $N$-fold coincidence, the leading edge of the coincidence window is aligned with one detector click.
If all other required clicks occur within the same coincidence window, a coincidence event is registered.
Notably, the same detection event can contribute to multiple coincidence events, as illustrated by fourfold events \#2 and \#3 in Fig.~\ref{Fig:S-coincidence}a.
This logic imposes different temporal constraints for different coincidence orders.
For fourfold coincidences relevant to HOM visibility measurement, all four photons must fall within the same coincidence window, whose width determines both the idler detection window $T$ and the signal detection window $T'=T$. 
For twofold coincidences, however, either photon may arrive first and define the leading edge of the coincidence window. 
Conditioned on an idler, any signal photon arriving within the interval $[-T,T]$ can therefore contribute to a twofold coincidence. 
This corresponds to a signal detection window $T'=2T$, as illustrated in the right panel of Fig.~\ref{Fig:S-coincidence}a, for heralding-efficiency extraction.

\added{
The standard single-window coincidence logic constrains the idler- and signal-photon detection-window widths, \(T\) and \(T'\), thereby introducing the purity--efficiency trade-off discussed in the main text. 
In contrast, the double-window coincidence logic illustrated in Fig.~\ref{Fig:S-coincidence}b allows \(T\) and \(T'\) to be configured independently. 
The detection channels are first assigned as idler channels (ch1 and ch4) and signal channels (ch2 and ch3). 
A fourfold event is accepted when the two idler clicks occur within an idler-photon detection-window of width \(T\), thereby heralding two signal photons incident on the HOM interferometer, and the two signal clicks both fall within a signal-photon detection-window of width \(T'\) centered at the mean time of the two idler clicks. 
In our experiment, this double-window coincidence logic is implemented by post-processing the raw time tags recorded.}

\section{Performance comparison}

Table~\ref{tab:S1} compares our results with state-of-the-art integrated-photonic HOM interference experiments using both CW and pulsed pumping.

\begin{table*}[h!]
\caption{\added{
\textbf{Performance comparison with state-of-the-art photonic integrated circuits.} 
The comparison includes representative demonstrations using both CW and pulsed pumps.
For microresonator sources, the photon spectral FWHM is estimated as $0.64\kappa/2\pi$ using the reported resonance linewidth; 
non-resonant broadband sources are labeled ``Broadband''.}
}
\centering
\begin{ruledtabular}
\begin{tabular}{lccccc}
\multirow{2}*{Reference} & \multirow{2}*{Platform} & \multirow{2}*{Pump type}& Linewidth & HOM interference & Fourfold count  \\
 & & & (GHz) & visibility & rate (Hz) \\[1.5pt]
\colrule\noalign{\vspace{0.8ex}}
This work & Si$_3$N$_4$ & CW & 0.1 & 0.992(8) & 4.5(3) \\[1.6pt]
This work & Si$_3$N$_4$ & CW & 0.1 & 0.942(12) & 12.2(6) \\[1.6pt]
\colrule\noalign{\vspace{0.8ex}}
Samara \emph{et al.} \cite{Samara:21} & Si$_3$N$_4$ & CW & 0.3 & 0.932 & 0.13 \\[1.6pt]
PsiQuantum team \cite{Alexander:25} & Si & Pulse & $\sim$2 & 0.995(3) & 0.2 \\[1.6pt]
Llewellyn \emph{et al.} \cite{Llewellyn:20} & Si & Pulse & 5 & 0.72 & $\sim$0.5 \\[1.6pt]
Bao \emph{et al.} \cite{Bao:23}& Si & Pulse & Broadband & 0.78 & 0.02 \\[1.6pt]
Adcock \emph{et al.} \cite{Adcock:19} & Si & Pulse & Broadband & 0.82 & 0.02 \\[1.6pt]
Paesani \emph{et al.} \cite{Paesani:20} & Si & Pulse & Broadband & 0.96 & 0.006 \\[1.6pt]
\end{tabular}
\end{ruledtabular}
\label{tab:S1}
\end{table*}

\section*{Supplementary References}
\bibliographystyle{apsrev4-1}
\bibliography{bibliography, bib_add}